\documentclass[accepted]{melba}

\usepackage{mwe} 

\usepackage{amsmath,amsfonts}

\usepackage{siunitx}
\usepackage{booktabs}
\usepackage{needspace}

\melbaid{2025:034}  
\doi{10.59275/j.melba.2025-gc8c}
\melbaauthors{Hansen et al.}  
\email{mattias.heinrich@uni-luebeck.de}
\volume{3}
\firstpageno{775}  
\melbayear{2025}  
\datesubmitted{2025-08-15}  
\datepublished{2025-12-09}  

\ShortHeadings{Learn2Reg 2024}{Hansen et al.}

\title{Learn2Reg 2024: New Benchmark Datasets Driving Progress on New Challenges}

\author{
    \firstname Lasse \surname Hansen\aff{1}\orcid{0000-0003-3963-7052}
    \firstname Wiebke \surname Heyer\aff{2}
    \firstname Christoph \surname Großbröhmer\aff{2}
    \firstname Frederic \surname Madesta\aff{3,4}
    \firstname Thilo \surname Sentker\aff{3,4}
    \firstname Wang \surname Jiazheng\aff{5}
    \firstname Yuxi \surname Zhang\aff{5}
    \firstname Hang \surname Zhang\aff{6}
    \firstname Min \surname Liu\aff{5}
    \firstname Junyi \surname Wang\aff{7}
    \firstname Xi \surname Zhu\aff{7}
    \firstname Yuhua \surname Li\aff{8}
    \firstname Liwen \surname Wang\aff{8}
    \firstname Daniil \surname Morozov\aff{9,10}
    \firstname Nazim \surname Haouchine\aff{9}
    \firstname Joel \surname Honkamaa\aff{11}
    \firstname Pekka \surname Marttinen\aff{11}
    \firstname Yichao \surname Zhou\aff{12}
    \firstname Zuopeng \surname Tan\aff{12}
    \firstname Zhuoyuan \surname Wang\aff{13}
    \firstname Yi \surname Wang\aff{13}
    \firstname Hongchao \surname Zhou\aff{14}
    \firstname Shunbo \surname Hu\aff{14}
    \firstname Yi \surname Zhang\aff{15}
    \firstname Qian \surname Tao\aff{15}
    \firstname Lukas \surname Förner\aff{16}
    \firstname Thomas \surname Wendler\aff{16}
    \firstname Bailiang \surname Jian\aff{10, 17}
    \firstname Christian \surname Wachinger\aff{10, 17}
    \firstname Jin \surname Kim\aff{18}
    \firstname Dan \surname Ruan\aff{18}
    \firstname Marek \surname Wodzinski\aff{19,20}
    \firstname Henning \surname Müller\aff{21,22}
    \firstname Tony C.W. \surname Mok\aff{23}
    \firstname Xi \surname Jia\aff{24}
    \firstname Jinming \surname Duan\aff{24}
    \firstname Mikael \surname Brudfors\aff{25}
    \firstname Seyed-Ahmad \surname Ahmadi\aff{26}
    \firstname Yunzheng \surname Zhu\aff{27}
    \firstname William \surname Hsu\aff{27}
    \firstname Tina \surname Kapur\aff{9}
    \firstname William M. \surname Wells\aff{9}
    \firstname Alexandra \surname Golby\aff{9}
    \firstname Aaron \surname Carass\aff{28}
    \firstname Harrison \surname Bai\aff{28}
    \firstname Yihao \surname Liu\aff{29}
    \firstname Perrine \surname Paul-Gilloteaux\aff{30}
    \firstname Joakim \surname Lindblad\aff{31}
    \firstname Nata\v{s}a \surname Sladoje\aff{31}
	\firstname Andreas \surname Walter\aff{32}
    \firstname Junyu \surname Chen\aff{28}
    \firstname Reuben \surname Dorent\aff{9,33}
    \firstname Alessa \surname Hering\aff{34,*}\orcid{0000-0002-7602-803X}
    \firstname Mattias P. \surname Heinrich\aff{2,*}\orcid{0000-0002-7489-1972}
}
\affiliations{
	\num 1 \addr EchoScout GmbH, Lübeck, DE
    \num 2 \addr Institute of Medical Informatics, Universität zu Lübeck, Lübeck, DE
    \num 3 \addr Institute of Applied Medical Informatics, University Medical Center Hamburg-Eppendorf, Hamburg, DE
    \num 4 \addr Institute of Computational Neuroscience, University Medical Center Hamburg-Eppendorf, Hamburg, DE
    \num 5 \addr School of Artificial Intelligence and Robotics, Hunan University, Changsha, CN
    \num 6 \addr Cornell University, New York, US
    \num 7 \addr University of Electronic Science and Technology of China, Chengdu, CN
    \num 8 \addr Mechanical Engineering Department, Tianjin University, Tianjin, CN
    \num 9 \addr Harvard Medical School, Brigham and Women's Hospital, Boston, US
    \num 10 \addr Technical University of Munich, Munich, DE
    \num 11 \addr Aalto University, Espoo, FI
    \num 12 \addr Canon Medical Systems (China) Co., Ltd., Beijing, CN
    \num 13 \addr Smart Medical Imaging, Learning and Engineering (SMLE) Lab, School of Biomedical Engineering, Shenzhen University Medical School, Shenzhen University, Shenzhen, CN
    \num 14 \addr School of Information Science and Engineering, Linyi University, Linyi, CN
    \num 15 \addr Department of Imaging Physics, Delft University of Technology, Delft, NL
    \num 16 \addr Clinical Computational Medical Imaging Research, Department of Diagnostic and Interventional Radiology and Neuroradiology, University Hospital Augsburg, Augsburg, DE
    \num 17 \addr TUM Klinikum Rechts der Isar, Munich, DE
    \num 18 \addr University of California, Los Angeles, US
    \num 19 \addr AGH University of Krakow, Krakow, PL
    \num 20 \addr Sano Centre for Computational Medicine, Krakow, PL
    \num 21 \addr Institute of Informatics, University of Applied Sciences Western Switzerland, Sierre, CH
    \num 22 \addr Medical Faculty, University of Geneva, Geneva,CH
    \num 23 \addr Hong Kong University of Science and Technology, Hong Kong, HK
    \num 24 \addr School of Computer Science, University of Birmingham, Birmingham, UK
    \num 25 \addr NVIDIA Ltd., London, UK
    \num 26 \addr NVIDIA GmbH, Munich, DE
    \num 27 \addr Medical \& Imaging Informatics, Department of Radiological Science, University of California, Los Angeles, US
    \num 28 \addr Johns Hopkins University, Baltimore, US
    \num 29 \addr Vanderbilt University, Nashville, US
    \num 30 \addr Nantes Université, CHU Nantes, CNRS, Inserm, BioCore, US16, SFR Bonamy, F 44000 Nantes, FR
    \num 31 \addr Uppsala Universitet, Uppsala, SE
    \num 32 \addr Aalen University, Aalen, DE
    \num 33 \addr National Institute for Research in Computer Science and Control, Paris, FR
    \num 34 \addr Radboud University Medical Center, Nijmegen, NL \\
    \num * \addr These authors contributed equally to this work and share senior authorship.
}

\abstract{
	Medical image registration is critical for clinical applications, and fair benchmarking of different methods is essential for monitoring ongoing progress in the field. To date, the Learn2Reg 2020-2023 challenges have released several complementary datasets and established metrics for evaluations. Building on this foundation, the 2024 edition expands the challenge's scope to cover a wider range of registration scenarios, particularly in terms of modality diversity and task complexity, by introducing three new tasks, including large-scale multi-modal registration and unsupervised inter-subject brain registration, as well as the first microscopy-focused benchmark within Learn2Reg. The new datasets also inspired new method developments, including invertibility constraints, pyramid features, keypoints alignment and instance optimisation.
	Visit Learn2Reg at~\url{https://learn2reg.grand-challenge.org}.}


\begin{document}

\twocolumn[\maketitle]

\section{Introduction}
    \enluminure{T}{his} paper serves as the official summary and documentation of the  Learn2Reg 2024 challenge, held in conjunction with MICCAI 2024. As an  overview paper, it introduces three new benchmark datasets addressing  previously underrepresented registration scenarios, documents the methods  developed by participating teams, and provides a comparative analysis of  their performance across tasks. While this paper offers a comprehensive  cross-task perspective on methods and results, individual task organizers  may publish separate, in-depth analyses focusing on specific research  questions, extended evaluations, or foundational model development. Our  primary contributions in this challenge overview include: (1) establishing  new publicly available benchmarks for multimodal intraoperative  registration, large-scale unsupervised brain alignment, and histopathology  microscopy registration; (2) documenting and comparing diverse  methodological approaches across these tasks; and (3) identifying current  limitations and future directions for medical image registration research  through cross-task insights.
    
	The motivation behind the Learn2Reg 2024 benchmark stems from the imperative need to enhance the robustness and versatility of medical image registration methodologies. Earlier challenges revealed limitations in dataset scale, modality diversity, and task complexity, highlighting the need for more comprehensive benchmarks. The 2024 edition addresses these gaps by introducing new tasks that include large-scale multi-modal and cross-domain registration, encouraging the development of more generalizable and adaptive algorithms.

    One of the key new tasks is ReMIND2Reg, which focuses on the multimodal alignment of intraoperative ultrasound (iUS) and MRI for brain tumor surgery. Compared to the earlier CuRIOUS2020 task, ReMIND2Reg provides a much larger dataset, increasing from a few dozen to several hundred scan pairs. Moreover, while CuRIOUS2020 targeted the alignment of pre-dura opening 3D iUS to pre-operative MRI, ReMIND2Reg tackles the more challenging alignment of post-resection iUS with pre-operative MRI. This setting reflects real-world surgical challenges and introduces greater anatomical variability, thereby promoting the development of more precise and robust registration methods for neurosurgical applications.
    
    The second task introduced is LUMIR, which focuses on large-scale inter-subject brain alignment in T1-weighted MRI and is designed without label supervision. It directly addresses a key limitation of earlier datasets such as OASIS, where the reliance on automatically generated labels risked algorithm overfitting and reduced generalizability. By removing label supervision, LUMIR encourages the development of more generalized and robust models that can better adapt to variations in brain structures.

    \begin{figure}[h]
		\centering
		\includegraphics[width=1\linewidth]{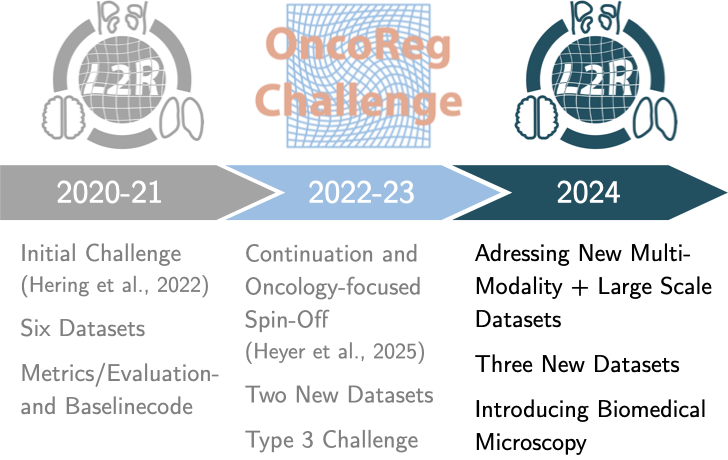}
		\caption{Learn2Reg 2024 continues the Learn2Reg Challenge series, which began in 2020, by introducing three new diverse datasets and expanding the challenge’s scope to cover a wider range of biomedical imaging applications.}
	\end{figure}
    
    Lastly, The COMULISglobe task introduces the first Learn2Reg benchmark focused on histopathology microscopy, marking a major expansion beyond radiology-based tasks. It represents a significant expansion of Learn2Reg's scope, catering to a wider array of clinical needs and fostering the advancement of cutting-edge techniques such as invertibility constraints, pyramid features, and keypoints alignment. These developments are crucial for achieving superior registration performance and advancing medical imaging applications. For example, they enable informative multimodal characterization of complex interactions in the tumor microenvironment, leading to improved personalized medicine strategies through accurate tracking of morphological changes over time.
    
\section{Related Works}
\subsection{Medical Image Registration Challenges}
    Several notable medical image registration challenges have significantly advanced the field. The EMPIRE10 challenge focused on lung CT registration, offering comprehensive evaluations including landmark distances and Jacobian determinants \citep{murphy2011evaluation}. The CuRIOUS challenge addressed intra-operative ultrasound and MRI registration \citep{xiao2019evaluation}, while ANHIR specialized in histological image registration \citep{borovec2020anhir}, each attracting multiple participating teams and employing diverse evaluation metrics. The Continuous Registration Challenge, co-hosted with WBIR 2018, aimed to unify multiple registration tasks but was limited by its integration requirements, specifically the mandatory use of the SuperElastix C++ API framework \citep{marstal2019continuous}. More recently, the ACROBAT challenges in 2022 and 2023 addressed whole-slide image registration for breast cancer tissue, utilizing large datasets and providing valuable insights into algorithm performance \citep{weitz2024acrobat}. These challenges have collectively advanced the development and benchmarking of medical image registration algorithms, each contributing unique datasets and frameworks that have influenced Learn2Reg tasks and evaluation.
\subsection{Learn2Reg to Date}
	Since its inception, the Learn2Reg challenge has served as a key platform for advancing medical image registration through open benchmarking and community collaboration. Over multiple editions, the challenge has brought together researchers worldwide to tackle diverse registration problems across various imaging modalities and anatomical sites.
    
    The Learn2Reg series began with the Learn2Reg Tutorial at MICCAI 2019 in Shenzhen, which introduced the community to emerging deep learning approaches for medical image registration~\citep{learn2reg2019tutorial}. This foundational event set the stage for the challenge's collaborative and multidisciplinary focus.
    
    The 2020 and 2021 editions of Learn2Reg were the initial official challenges, introducing six datasets covering multiple anatomies—including brain, abdomen, and thorax—and imaging modalities such as MRI, CT, and ultrasound. Both editions were held as community challenges with workshops at the respective MICCAI conferences, following a Type 1 challenge format with training images and labels provided, while test labels remained withheld for evaluation. These editions introduced the challenge's open-source evaluation and ranking code, as well as baseline methods, enabling standardized and reproducible benchmarking. Together, they attracted over 20 international teams submitting more than 65 methods, offering comprehensive insights into the performance of conventional and deep-learning registration approaches across multiple clinically relevant tasks~\citep{hering2022learn2reg}.
    
    In 2022 and 2023, Learn2Reg held two further challenge workshops at MICCAI, focusing on continuation and expansion of the initial benchmarks. These editions introduced the NLST dataset (CT Lung Screening Registration Data) and pushed towards Type 3 challenges for selected tasks, where both training and test images, along with their labels, remain hidden during evaluation to better simulate real-world conditions.
    
    In parallel, the OncoReg spin-off challenge was launched at MICCAI 2023, introducing the ThoraxCBCT dataset with longitudinal lung CBCT and FBCT scans from clinical settings. OncoReg targets oncological applications to advance AI-driven registration for monitoring lung cancer progression and improving radiotherapy planning~\citep{heyer2025oncoreg}.
    
    The 2024 edition continues this trajectory, expanding into new imaging domains including intraoperative multimodal imaging and microscopy, while addressing increasingly complex registration problems.

\section{Materials and Methods}
\subsection{Challenge Organisation}
    The Learn2Reg 2024 challenge was organised by a dedicated core team led by Mattias Heinrich (University of Lübeck), who coordinated all aspects of the event, including website maintenance, evaluation infrastructure, and the associated workshop. The challenge consisted of three tasks, each developed and maintained by a separate expert team: ReMIND2Reg (Harvard Medical School, led by Reuben Dorent), LUMIR (Johns Hopkins University, led by Junyu Chen), and COMULISglobe SHG/BF (COMULISglobe community, led by Andreas Walter, Hochschule Aalen, \citep{walter2021correlative, rudraiah2024correlated, walter2020correlated}). These tasks addressed diverse challenges in biomedical image registration and are described in detail in later sections.
    
     Returning to a type 2 challenge format, Learn2Reg 2024 followed a two-phase structure hosted on the Grand Challenge platform\footnote{\url{https://learn2reg.grand-challenge.org/}}. During the validation phase, participants were given access to training and validation data to develop and test their methods locally. In the subsequent test phase, participants submitted their final algorithms, which were evaluated on hidden test datasets. The evaluation was performed using a standardized, open-source framework\footnote{\url{https://github.com/MDL-UzL/L2R}}, ensuring fairness, reproducibility, and transparency. Performance was assessed using task-specific quantitative metrics, with results published on public leaderboards. The ranking scheme is based on the ranking scheme of the Medical Decathlon (statistical comparisons of method performance across metrics, aggregated into overall task scores) \citep{antonelli2022medical}.
    
    For the first time, Learn2Reg was part of the WBIR (Workshop on Biomedical Image Registration), which was officially aligned with MICCAI and held as an on-site half-day workshop at MICCAI 2024 in Marrakesh. Top-performing teams were invited to present their methods and findings during this event.

\subsection{Tasks}
    \begin{figure}[htbp!]
        \centering
        \includegraphics[width=1\columnwidth]{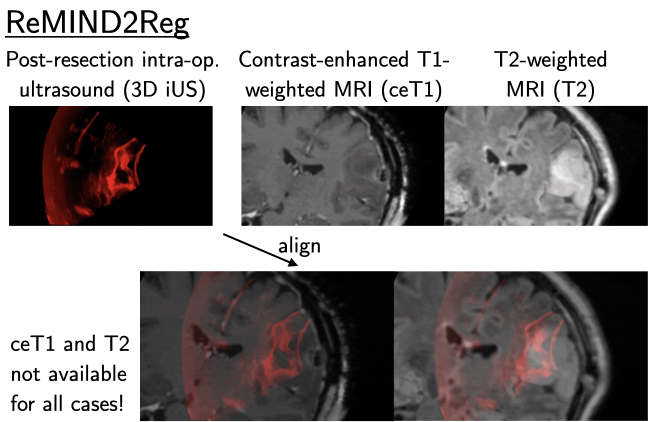}\\
        \vspace{0.5cm}
        \includegraphics[width=1\columnwidth]{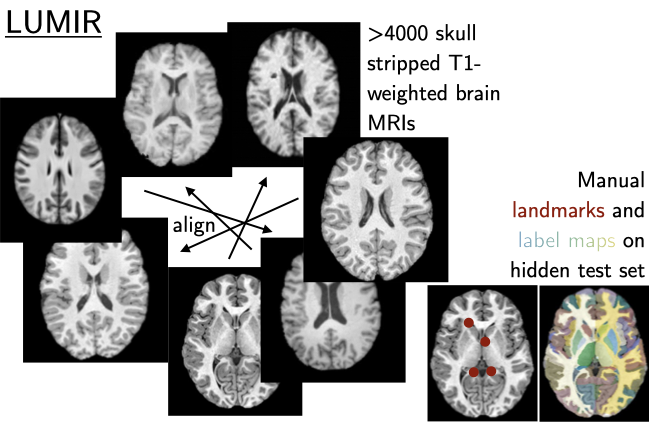}\\
        \vspace{0.5cm}
        \includegraphics[width=1\columnwidth]{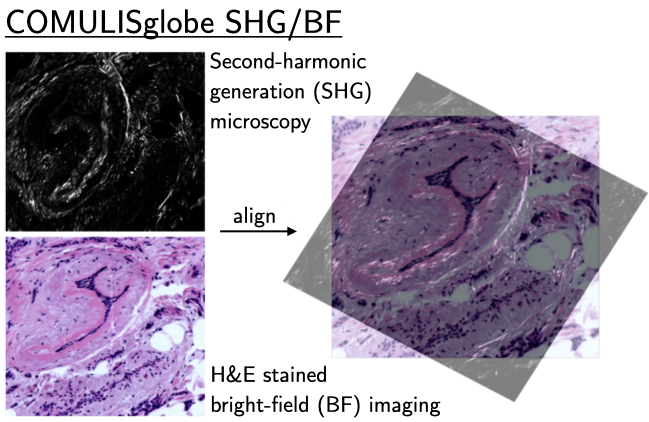}
        \caption{Learn2Reg 2024 tasks at a glance. This overview illustrates the diversity of the three new tasks in Learn2Reg 2024. ReMIND2Reg challenges participants with intraoperative ultrasound-to-MRI brain registration, addressing brain shift and flexible input modalities. LUMIR focuses on large-scale, inter-subject brain alignment without label supervision, using over 4,000 scans from 10 input sources. \mbox{COMULISglobe} SHG/BF introduces the first histology-based task in the Learn2Reg series, aligning complementary breast tissue modalities—second harmonic generation (SHG) and brightfield (BF) microscopy. All tasks include hidden test data and are evaluated using manual landmarks, label maps, or both.}
    \end{figure}
\subsubsection{ReMIND2Reg}
    ReMIND2Reg focuses on a clinically critical but technically complex task: registering preoperative brain MRI to intraoperative ultrasound images after tumor resection. Intraoperative ultrasound (iUS) offers real-time imaging capabilities during surgery but differs significantly in appearance and resolution from preoperative MRI, complicating direct multimodal registration. The surgical process itself induces brain shift and structural deformation, making static navigation systems inaccurate as surgery progresses.

    The challenge task is to register post-resection 3D iUS volumes to either contrast-enhanced T1-weighted (ceT1) or T2-weighted MRI, depending on availability for each case. Submitted methods must be able to handle missing modalities and account for large deformations, resections, and noise inherent in iUS data. The dataset is derived from the ReMIND collection \citep{juvekar2023remind} at Brigham and Women's Hospital. Specifically,  the ReMIND2Reg dataset includes 99 training cases with 99 3D iUS, 93 ceT1, and 62 T2 scans, 4 validation cases with 4 3D US, 4 ceT1, and 4 T2, and 10 test cases with 10 3D US, 10 ceT1 and 10 T2.
    
    All data have been resampled to 0.5 mm isotropic spacing and MRI modalities co-registered where applicable. The MRI sequences used include contrast-enhanced T1 and T2-SPACE, while ultrasound was acquired using tracked 2D sweeps reconstructed into 3D volumes. Manual landmark annotations created by multiple expert raters serve as the ground truth for validation and testing. For evaluation, the  target registration error (TRE) were computed. Participants submit Docker containers for test evaluation. 
    
    Full data and acquisition protocols are described in \citep{juvekar2023remind}. The dataset is shared under a CC BY 4.0 license and is available via TCIA\footnote{\url{https://doi.org/10.1101/2023.09.14.23295596}}.
    
\subsubsection{LUMIR}
    LUMIR (Large Scale Unsupervised Brain MRI Image Registration) addresses the alignment of T1-weighted brain MRIs from different subjects in an unsupervised setting. The task builds on previous editions of the Learn2Reg challenge and aims to overcome limitations observed when training with anatomical label maps, which can result in overfitting to labels and non-smooth transformations. By omitting label supervision during training, the challenge promotes the development of models that generalize better and produce realistic, diffeomorphic deformations. This shift also allows fair comparison with classical optimization-based registration algorithms.

    Participants are given access to a preprocessed large-scale dataset of 3,384 training images compiled from ten public neuroimaging datasets (OpenBHB) and an additional set from the AFIDs-OASIS dataset. Each MRI has been resampled to 1 mm isotropic resolution and standardized to a shape of 160$\times$224$\times$192 voxels. Label maps and anatomical landmarks are retained for evaluation only and are not made available during training. Validation data includes 40 images from OpenBHB and 10 with manually placed fiducials from AFIDs-OASIS. Testing is based on 590 hidden images, with evaluation pairs defined by the organizers. Among them, 20 images are from AFIDs-OASIS with manually annotated landmarks, and 110 images were annotated in-house by a team of physicians led by an experienced radiologist (H. Bai) following the AFIDs protocol. The remaining test images are accompanied by label maps of 133 anatomical structures automatically generated using the SLANT method~\citep{huo20193d}.
    
    Participants are required to submit deformation fields aligning specified image pairs. The performance is assessed via Dice similarity and Hausdorff-95 distance on segmentation labels, target registration error on anatomical landmarks, and the fraction of non-diffeomorphic volumes (NDV)~\citep{liu2024jacobian} to measure deformation plausibility. Baseline code and evaluation scripts are provided online\footnote{\url{https://github.com/JHU-MedImage-Reg/LUMIR_L2R}}. An introductory video is also available\footnote{\url{https://cloud.imi.uni-luebeck.de/s/tm9DEFiDa9XH35k/download/Learn2Reg_LUMIR_JChen.mp4}}.
    
    This section documents LUMIR within the Learn2Reg 2024 challenge framework. As a multi-task overview, we focus on describing the task design, documenting participating methods, presenting comparative rankings, and identifying cross-task insights, particularly regarding the impact of dataset scale on classical versus deep learning approaches. The value of including LUMIR lies in enabling direct comparison with ReMIND2Reg and COMULISglobe, and providing a complete record of all Learn2Reg 2024 tasks. A complementary in-depth analysis with extensive ablation studies, zero-shot generalization experiments, and foundational model evaluation is provided in a dedicated paper \citep{chen2025LUMIRchallengepathwayfoundational}. Relevant datasets and prior methods are further detailed in \citep{dufumier2022openbhb, taha2023magnetic, marcus2007oasis, liu2024jacobian}.
\subsubsection{COMULISglobe SHG/BF}
    COMULISglobe SHG/BF addresses the registration of histo\-pathology images acquired using two fundamentally different microscopy techniques: second-harmonic generation (SHG) and bright-field (BF) imaging of H\&E-stained tissue. SHG provides high-resolution structural information of collagen fibers without the need for staining, while BF microscopy reveals cellular and tissue structure via conventional staining. Registering these modalities is essential for correlating structural and pathological insights and is particularly valuable in cancer research where collagen architecture plays a key role.

    This task involves registering co-located SHG and BF images of breast and pancreatic cancer tissue sections. Tissue was sectioned at \SI{5}{\micro\metre} thickness, stained, and mounted before imaging. SHG images were acquired at three axial depths and processed with maximum-intensity projection to compensate for focal plane inconsistencies. BF imaging was performed using commercial and custom-built platforms at 40x magnification.
    
    Participants work with 156 training samples, 10 for validation, and 40 for testing. Manual landmark annotations are available for validation and test sets. The challenge evaluates registration accuracy using target registration error, and participants submit displacement fields as their output. This task is particularly relevant for enabling virtual staining and rapid multimodal analysis in digital pathology, where SHG microscopy is gaining traction as a label-free diagnostic alternative.
    
    Further details on the underlying imaging and dataset are available in \citep{keikhosravi2020intensity}. More information on the goals of the COMULISglobe initiative and the broader community can be found at the COMULISglobe website\footnote{\url{https://www.comulis.eu/}}.
\subsubsection{NLST}
    In addition to the three newly introduced tasks in Learn2Reg 2024, this paper also documents the ongoing NLST task, first launched in 2022 and expanded in 2023.

    The NLST (Thoracic Image Registration for Lung Cancer) task targets a highly impactful clinical application: aligning intra-patient thoracic CT scans to facilitate lung nodule tracking and support therapy planning. Accurate registration of baseline and follow-up low-dose CT scans is essential in lung cancer screening programs to assess the progression of pulmonary nodules and metastatic lesions. Manual identification of corresponding structures is laborious and error-prone, particularly in the lung where many anatomical features appear similar. Deformable registration algorithms can significantly streamline this process and enable automated propagation of findings across time points.

    Beyond longitudinal screening, the task also includes a radiotherapy sub-cohort with 4DCT scans, where precise alignment is needed for treatment planning and the mapping of structures such as the trachea, esophagus, spinal cord, and heart. These organs-at-risk must be accurately registered to intraoperative cone-beam CT scans to ensure safe and effective dose delivery.

    The challenge consists of two phases. In Phase 1, participants developed and tested algorithms locally using 110 training and 22 validation cases, all sourced from the National Lung Screening Trial (NLST) \citep{national2011reduced} \footnote{\url{https://wiki.cancerimagingarchive.net/display/NLST/National+Lung+Screening+Trial}}. Each case consists of a pair of thoracic CT scans acquired either longitudinally (screening cohort) or across different respiratory phases (radiotherapy cohort). Automatic keypoint correspondences \citep{heinrich2015estimating} and lung masks are provided to support model development. In Phase 2, the top-performing teams were invited to submit training Docker containers for remote execution on a hidden, extended dataset comprising 1300 training, over 100 validation, and more than 200 test cases. This second part of the dataset includes additional annotations not publicly released.

    The evaluation framework measures registration performance using multiple metrics: target registration error (TRE) on lung nodule centers and manually annotated geometric landmarks (e.g. at bifurcations), deformation plausibility via the standard deviation of log Jacobian determinants, and runtime. Robustness is also assessed by considering the 30\% worst-performing cases per submission. 

    All data are preprocessed to standardized voxel spacing and affine-aligned. The CT data originate from over 30 U.S. medical centers participating in the NLST and from NSCLC radiotherapy patients with fan-beam 4DCT scans. Segmentations were created using CIRRUS lung screening software, with nodule locations derived from the center of gravity of annotated lesions. The dataset reflects diverse demographics and scanning equipment, adding to its clinical realism.

    The NLST task exemplifies the challenges of real-world thoracic image registration and is designed to evaluate both accuracy and generalisability of methods under clinically relevant constraints. 

\section{Results and Discussions}
    This section presents the results and key insights from each of the three Learn2Reg 2024 tasks as well as from NLST. For each task, we summarize the test setting, describe the participating teams and their approaches, highlight the top-performing methods, and discuss relevant findings, limitations, and future directions. The final subsection offers a cross-task analysis, including observations on team participation across tasks and overall challenge outcomes.
\subsection{Tasks}
\subsubsection{ReMIND2Reg}
    \begin{figure}[b!]
        \renewcommand{\arraystretch}{0.8}
        \centering
        \begin{tabular*}{\linewidth}{@{\extracolsep{\fill}} c l c c c}
        \toprule
        \# & Team/Method & TRE & TRE30 & Rank \\
        \midrule
        1 & niftyreg*           & 2.87  & 4.54  & 0.962 \\
        2 & junyi-wang          & 4.42  & 6.01  & 0.805 \\
        3 & VROC                & 3.63  & 6.24  & 0.661 \\
        3 & next\_gen\_nn       & 3.71  & 6.49  & 0.661 \\
        4 & IGRTS               & 4.94  & 7.42  & 0.325 \\
        5 & SamPLe              & 14.73 & 17.51 & 0.100 \\
        \midrule
          & Initial             & 4.81  & 7.50  &       \\
        \bottomrule
\end{tabular*}
        \includegraphics[width=1\linewidth]{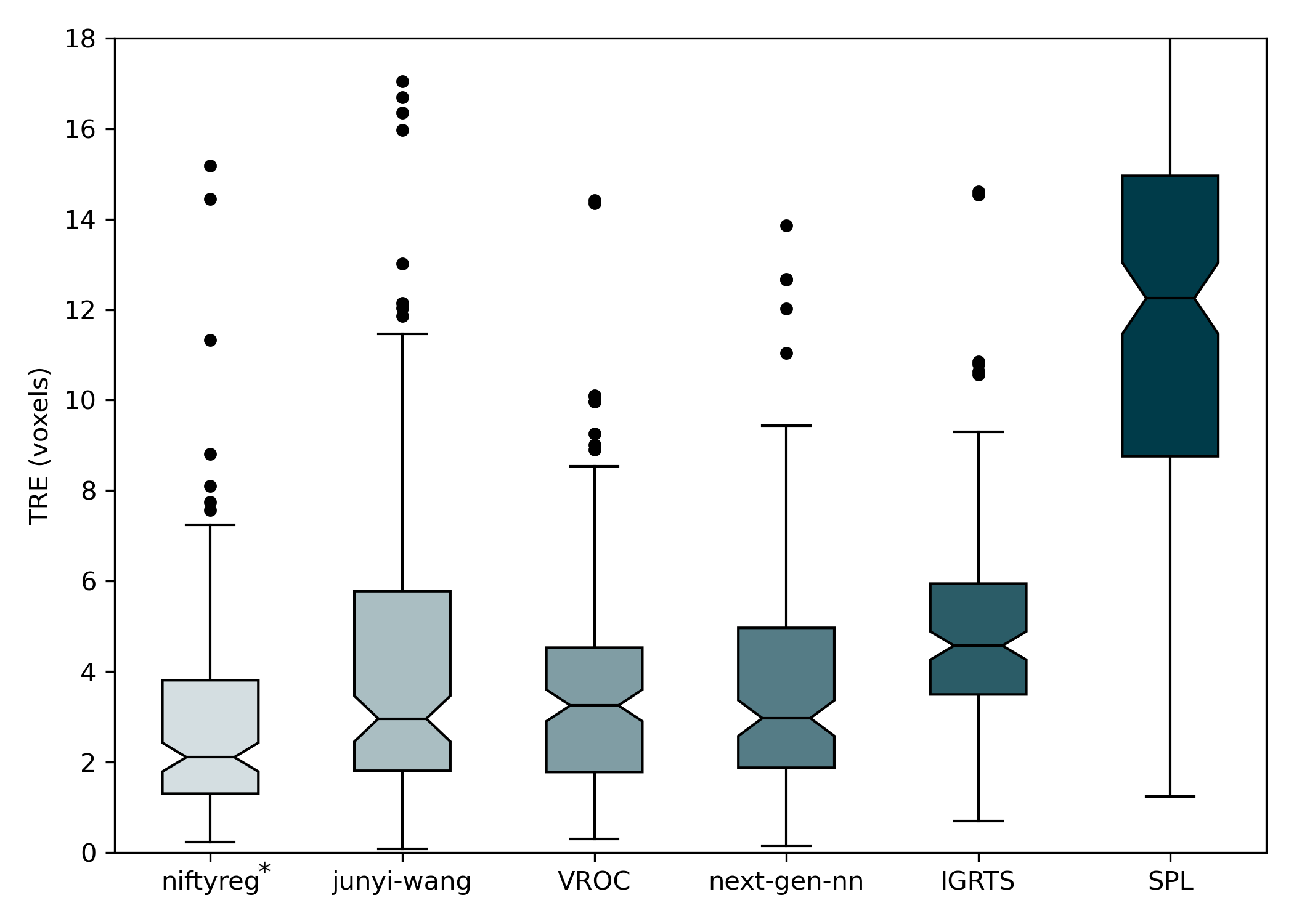}
        \caption{Results for the ReMIND2Reg task. Target Registration Error (TRE) on manual landmarks and TRE30 (the TRE computed on the 30\% of landmarks with the largest initial errors) are reported for all methods. * Baseline method provided by organizers.}
    \end{figure}
    During the validation phase of the ReMIND2Reg task, a total of thirteen teams submitted results, along with two baseline methods provided by the challenge organizers. The task's complexity was reflected in the exceptionally high number of validation entries, with 253 submissions indicating that teams required extensive iterative optimization to handle the challenging multimodal ultrasound-to-MRI registration problem. In the test phase, participation reduced to five teams plus one baseline method. This reduction highlights the technical difficulty of registering post-resection intraoperative ultrasound to preoperative MRI, particularly given the presence of brain shift and tissue deformation. As a baseline, niftyreg was provided for the ReMIND2Reg test phase. The submitting teams included junyi-wang, VROC, next-gen-nn, IGRTS, and SamPLe. 
    
    \textbf{niftyreg}, which employed the NiftyReg open-source software package\footnote{https://github.com/KCL-BMEIS/niftyreg/wiki} implementing a symmetric block-matching registration approach. This method used iterative block-matching to establish point correspondences between ultrasound and MRI images, followed by least trimmed squares (LTS) regression to determine transformation parameters. To identify informative regions, the algorithm divided images into uniform blocks and selected those with the highest intensity variance. Block similarity was assessed using normalized cross-correlation, and matching was performed based on these scores. To ensure inverse consistency, the method averaged directional transformation matrices at each iteration, maintaining symmetry throughout the registration process \citep{drobny2018registration}.
    
    \textbf{junyi-wang} developed a coarse-to-fine pipeline combining imaging style transfer with traditional registration techniques. Their approach used 3D CycleGAN to generate synthetic T1-style images from ultrasound data, enabling unified signal distributions across modalities. The registration process employed hierarchical block matching via NiftyReg for affine transformation estimation, followed by a pretrained SynthMorph model for local deformation inference around resection areas. \textbf{VROC} employed its fully differentiable variational registration framework \citep{werner2014estimation, ehrhardt2015variational}\footnote{\url{https://github.com/IPMI-ICNS-UKE/vroc}}, combining task-specific preprocessing with rigid registration optimization. For the ReMIND2Reg task, VROC applied Gaussian smoothing and threshold-based clipping of ultrasound intensity values during preprocessing. The registration pipeline performed initial vector field estimation via global normalized cross-correlation (NCC) loss optimization, followed by fine-tuning using normalized gradient fields (NGF) loss optimization to handle the challenging multimodal ultrasound-to-MRI registration scenario. 
    
    \textbf{next-gen-nn} addressed the ReMIND2Reg challenge using an unsupervised multimodal registration method based on Multilevel Correlation Balanced Optimization (MCBO), building upon the ConvexAdam framework \citep{siebert2024convexadam}. The approach was specifically designed to handle two key difficulties: the absence of supervised labels and the presence of large deformations between pre-operative MRI and intra-operative ultrasound images. For paired multimodal moving images (ceT1 and T2), the method adopted maximum fusion strategies to optimize registration performance across the challenging modality gap. \textbf{IGRTS} developed a topological higher-order Markov Random Field approach for deformable registration. Their method utilized multiscale multimodal Gabor attribute vectors to characterize MRI and ultrasound images, with the 3D deformation field's topological framework constructed through Jacobian determinants derived from higher-order cliques. The optimization employed a multi-resolution QPBO (Quadratic Pseudo-Boolean Optimization) strategy to handle the complex higher-order MRF energy function.
    
    \textbf{SamPLe} implemented a cross-modal feature matching approach using 3D keypoint descriptors specifically designed for MR-ultrasound registration. Their method trained a Siamese convolutional neural network on 3D patches extracted at keypoint locations using supervised contrastive learning with triplet loss. The approach detected keypoints using 3D SIFT, enforced repeatability across modalities, and learned cross-modal discriminant features robust to ultrasound texture changes and speckle noise. Final registration was performed by solving a least-squares problem using matched keypoint correspondences. \\

    The results demonstrate the substantial difficulty of the ReMIND2Reg task, with most methods showing modest improvements over the initial misalignment (TRE: 4.81 mm). Notably, the classical optimization-based baseline niftyreg achieved the best overall performance with a TRE of 2.87 mm and TRE30 of 4.54 mm, ranking first among all submissions. This finding highlights that well-tuned classical methods can still outperform learning-based approaches for challenging multimodal registration tasks with limited training data.
    Among the learning-based submissions, junyi-wang ranked second with a TRE of 4.42 mm, followed by VROC and next-gen-nn tied for third place with TREs of 3.63 mm and 3.71 mm respectively. IGRTS ranked fourth with a TRE of 4.94 mm, while SamPLe showed the poorest performance with a TRE of 14.73 mm, indicating substantial registration failures.
    The wide variation in performance across methods (TRE ranging from 2.87 to 14.73 mm) suggests that robustness remains a significant concern. Several methods achieved TREs that were only marginally better than, or in some cases worse than, the initial misalignment, indicating that the task remains largely unsolved for many approaches. The presence of outliers and failure cases across most methods emphasizes the need for more robust algorithms capable of handling the substantial anatomical deformations, tissue resection, and multimodal appearance differences inherent in this clinical scenario.
    The strong performance of the classical niftyreg baseline suggests that the explicit modeling of correspondences through block-matching and the iterative refinement process may be more effective than end-to-end learning approaches for this particular challenge. The substantial drop in participation from validation (13 teams) to test phase (5 teams) reflects the technical difficulty of the task. This pattern differs from other tasks in the challenge and underscores the need for continued research into multimodal intraoperative registration methods that can reliably handle brain shift and tissue deformation in clinical settings.
    Future work should focus on improving robustness through better handling of outliers, incorporating anatomical priors specific to brain tumor surgery, and developing hybrid approaches that combine the strengths of classical optimization with learning-based feature extraction and similarity measures.
    
\subsubsection{LUMIR}
    \begin{figure*}[t!]
        \renewcommand{\arraystretch}{0.8}
        \centering
        \begin{tabular*}{\linewidth}{@{\extracolsep{\fill}} c l c c c c c}
        \toprule
        \# & Team/Method & TRE & DSC & HD95 & NDV & Rank \\
        \midrule
        1 & honkamj                & 3.09  & 0.785 & 3.04 & 0.0025 & 0.814 \\
        2 & next\_gen\_nn          & 3.12  & 0.777 & 3.28 & 0.0001 & 0.781 \\
        3 & MadeForLife            & 3.07  & 0.778 & 3.29 & 0.0121 & 0.737 \\
        4 & zhuoyuanw210           & 3.14  & 0.773 & 3.23 & 0.0045 & 0.723 \\
        5 & LYU1                   & 3.13  & 0.778 & 3.25 & 0.0150 & 0.722 \\
        6 & DutchMasters           & 3.11  & 0.770 & 3.26 & 0.0030 & 0.702 \\
        7 & uniGradICONiso*        & 3.14  & 0.760 & 3.40 & 0.0002 & 0.668 \\
        8 & VFA*                   & 3.14  & 0.777 & 3.15 & 0.0704 & 0.667 \\
        9 & lukasf                 & 3.14  & 0.764 & 3.42 & 0.2761 & 0.561 \\
        10 & Bailiang              & 3.16  & 0.774 & 3.33 & 0.0222 & 0.526 \\
        11 & TransMorph*           & 3.14  & 0.762 & 3.46 & 0.3621 & 0.518 \\
        12 & TimH                  & 3.19  & 0.730 & 3.57 & 0.0000 & 0.487 \\
        13 & deedsBCV*             & 3.10  & 0.696 & 3.94 & 0.0002 & 0.423 \\
        14 & uniGradICON*          & 3.24  & 0.742 & 3.57 & 0.0001 & 0.402 \\
        15 & LoRA-FT               & 3.24  & 0.736 & 3.73 & 0.0033 & 0.384 \\
        16 & SynthMorph*           & 3.23  & 0.722 & 3.61 & 0.0000 & 0.361 \\
        17 & VROC                  & 3.23  & 0.760 & 3.63 & 0.0475 & 0.351 \\
        18 & ANTsSyN*              & 3.48  & 0.703 & 3.69 & 0.0000 & 0.265 \\
        19 & VoxelMorph*           & 3.53  & 0.714 & 4.07 & 1.2167 & 0.157 \\
        \midrule
         & Initial                 & 4.38  & 0.555 & 4.91 &        &       \\
        \bottomrule
        \end{tabular*}
        \includegraphics[width=\linewidth]{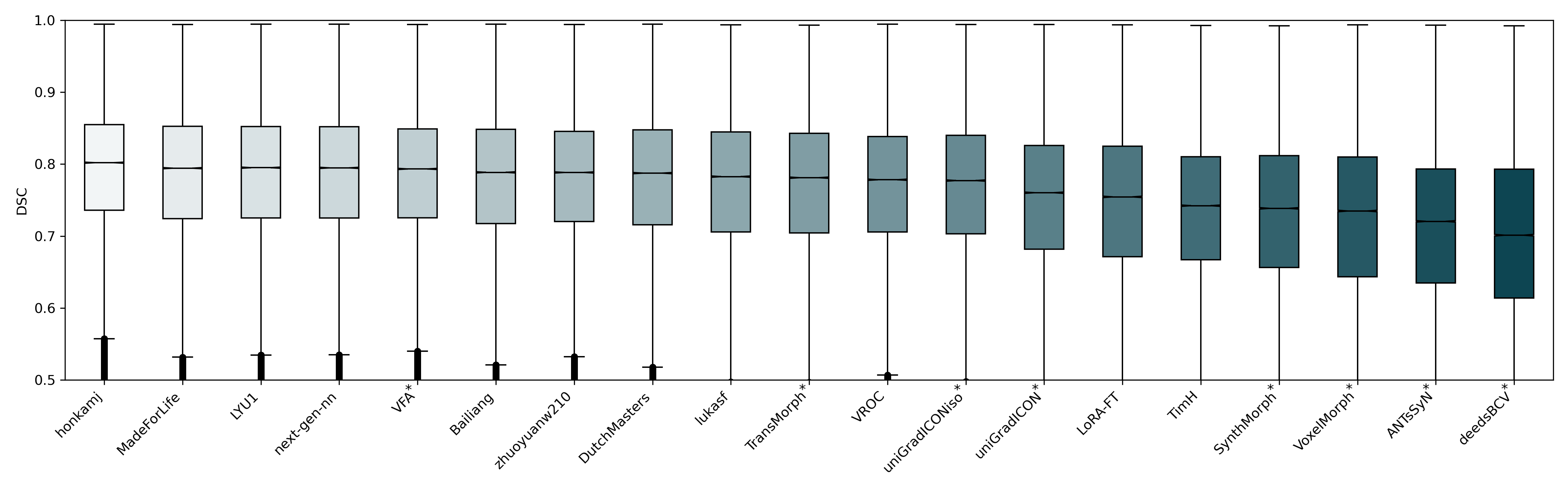}
        \caption{Results for the LUMIR task. The table presents final test set results for all participating teams and baseline methods. Reported metrics include Target Registration Error (TRE) on manual landmarks, Dice-Sørensen coefficient (DSC) and 95th percentile Hausdorff distance (HD95) on brain label maps, as well as the non-diffeomorphic volume (NDV) of displacement fields. The accompanying box plot provides a visual overview of DSC performance distribution across all methods. * Baseline method provided by organizers.}
    \end{figure*}
    During the validation phase of the LUMIR task, a total of eighteen teams submitted results, with four baseline methods provided by the challenge organizers. In the test phase, the number of submitting teams reduced to eleven, while the organizers expanded the baseline comparisons to include four additional methods, providing eight baselines in total. This comprehensive baseline coverage enabled thorough benchmarking across diverse registration approaches. Notably, LUMIR attracted strong participation. The high level of engagement reflects the strong community interest in large-scale, unsupervised brain MRI registration tasks.
    
    A distinctive aspect of the LUMIR challenge was its extensive post-challenge analysis, documented in detail in "Beyond the LUMIR challenge: The pathway to foundational registration models" \citep{chen2025LUMIRchallengepathwayfoundational}. This companion paper provides comprehensive validation with particular emphasis on foundational registration models, zero-shot evaluation, and out-of-domain generalization tasks. The present section summarizes the key findings and participating methods, while readers are referred to the companion paper for in-depth technical analysis and broader implications for developing generalizable brain MRI registration frameworks. The LUMIR challenge included eight baseline methods spanning classical optimization-based approaches and modern deep learning frameworks. \textbf{ANTsSyN} employed symmetric, time-varying diffeomorphic registration using cross-correlation within the space of diffeomorphic maps \citep{avants2008symmetric}. \textbf{deedsBCV} utilized discrete optimization with Markov random field strategies and minimum spanning trees to handle sliding motions efficiently \citep{heinrich2012mind}. \textbf{SynthMorph} implemented a contrast-agnostic learning strategy trained exclusively on synthetic data generated from label maps \citep{hoffmann2021synthmorph}. \textbf{TransMorph} incorporated Vision Transformer architecture with a three-stage Swin Transformer encoder for hierarchical feature extraction \citep{chen2022transmorph}. \textbf{uniGradICON} and \textbf{uniGradICONiso} represented foundation registration models using gradient inverse consistency regularization, with the latter including instance-specific optimization \citep{tian2024unigradicon}. \textbf{VFA} employed dual-stream encoders with vector field cross-attention decoders for pixel-level correspondence retrieval \citep{liu2024vector}. \textbf{VoxelMorph} served as the canonical U-Net-based learning framework for fast deformable registration \citep{balakrishnan2019voxelmorph}. The top-performing methods in LUMIR employed diverse architectural strategies, ranging from specialized ConvNet designs to foundation model adaptations.
    
    \textbf{Bailiang} utilized a dual-stream encoder architecture with registration-specific components, employing coarse-to-fine optimization pyramids and progressive registration to handle large deformations through iterative composition of deformation fields. Correlation layers were used to establish voxel-wise correspondences between moving and fixed images \citep{jian2024mamba}.
    
    \textbf{next-gen-nn} implemented a robust backbone network \citep{zhang2024memwarp} with several architectural enhancements, including co-attention mechanisms \citep{chen2021car}, large kernel convolutions, and increased convolutional channels in the deformation prediction stage. The method was further refined using image-guided bilateral filtering to improve registration accuracy.

    \textbf{honkamj} introduced a novel approach focusing on mathematical guarantees, implementing inverse consistency, symmetry, and topology preservation by construction rather than through loss terms. The method employed an auxiliary network formulation where $f(A,B) = u(A,B) \circ u(B,A)^{-1}$, ensuring cycle-consistency across a multi-resolution pipeline. Features were extracted separately before deformation computation, with symmetric halfway deformations composed at the highest resolution \citep{honkamaa2023sitreg}.

    \textbf{LoRA-FT} addressed the computational challenges of foundation model adaptation by applying Low-rank Adaptation (LoRA \citep{hu2022lora}) to uniGradICON foundation models \citep{tian2024unigradicon}. This approach decomposed weight updates into low-rank matrices, reducing trainable parameters by orders of magnitude while enabling rapid domain adaptation without catastrophic forgetting or requiring anatomical label supervision.

    \textbf{MadeForLife} employed the GroupMorph \citep{tan2024groupmorph} structure with dual encoders and specialized grouping update modules (GUMs). The method decomposed deformation fields into multiple subfields with different receptive fields, facilitated by grouping modules for feature correlation and contextual fusion modules for inter-group communication, enabling finer-grained optimization.
    
    \textbf{lukasf} refined a pre-trained TransMorph \citep{chen2022transmorph} model using the Fisher Adam (FAdam) optimizer \citep{hwang2024fadam} for improved convergence stability and incorporated gradient correlation in the similarity measure to enhance deformation smoothness and anatomical alignment.

    \textbf{LYU1} implemented a pyramid network architecture with shared encoder and auxiliary decoder components. The fusion pyramid decoder operated across five scales, using multi-scale feature fusion blocks (MSFB) to iteratively combine coarse deformation fields. The method progressively refined deformations by upsampling and applying fields to moving feature maps at each scale, removing redundant information through the MSFB processing.
    
    \textbf{TimH} employed the ConstrICON framework \citep{greer2023inverse} to achieve multistep inverse consistency through architectural design rather than penalty terms. The approach used three residual U-Nets in a recursive TwoStepConsistent (TSC) operator structure, parameterizing transformations via Lie groups. Stationary velocity fields were smoothed using average pooling, and selective Jacobian determinant regularization was applied to mitigate folding.
    
    \textbf{VROC} utilized a classical variational registration approach with task-specific preprocessing, including Gaussian smoothing and multi-Otsu thresholding to generate intensity-based label maps. The method performed initial deformable registration using normalized cross-correlation (NCC) forces on label maps, followed by Demons forces refinement on original intensity images for precise alignment \citep{werner2014estimation, ehrhardt2015variational}.
    
    \textbf{DutchMasters} introduced a novel instance optimization algorithm addressing multi-objective optimization challenges in registration. The method applied gradient projection techniques to handle conflicting updates between similarity and regularization objectives, projecting conflicting gradients into common spaces when cosine similarity indicated gradient conflicts, while preserving non-conflicting gradient updates \citep{zhang2024improving}.
    
    \textbf{zhuoyuanw210} employed a dual-stream pyramid encoder paired with a GPU-accelerated motion decomposition Transformer (ModeTv2) decoder \citep{wang2024modetv2}. The dual-stream architecture extracted hierarchical features from both moving and fixed images, generating deformations from coarse to fine across multiple scale levels.

    The results demonstrate strong performance across both learning-based and classical approaches, with honkamj achieving the highest overall rank (0.814), followed by next-gen-nn (0.781) and MadeForLife (0.737) completing the top three.
    Notably, the top-performing methods achieved remarkably similar performance across all metrics, with the leading submissions showing narrow ranges in Target Registration Error (honkamj: 3.09, MadeForLife: 3.07, next-gen-nn: 3.12), Dice scores (0.773-0.785), and Hausdorff distances (3.04-3.29). This tight clustering of results suggests that multiple architectural strategies can achieve comparable accuracy for unsupervised brain MRI registration when properly optimized.
    The LUMIR challenge highlighted the significant impact of dataset scale on method performance. With over 4,000 brain MRI scans available for training, a clear trend emerged favoring deep learning approaches over classical optimization-based methods. Non-deep learning methods including VROC (rank: 0.351), ANTsSyN (rank: 0.265), and deedsBCV (rank: 0.423) consistently ranked in the lower half of submissions, contrasting with other Learn2Reg 2024 tasks where classical methods remained competitive with learning-based approaches.
    Foundation model approaches showed mixed results. While uniGradICONiso (rank: 0.668) with instance optimization performed reasonably well, the base uniGradICON model (rank: 0.402) ranked lower, emphasizing the importance of task-specific adaptation strategies. LoRA-FT's parameter-efficient adaptation approach (rank: 0.384) demonstrated the potential for foundation model fine-tuning, though further optimization may be needed to compete with top-performing specialized architectures.
    The challenge successfully demonstrated the viability of unsupervised approaches for large-scale brain MRI registration, with all methods substantially outperforming the unregistered baseline (initial TRE: 4.38, DSC: 0.555). The removal of label supervision during training encouraged the development of more generalizable models while maintaining competitive registration accuracy, as evidenced by the consistently strong Dice scores ($>0.73$) among the top ten methods.

\subsubsection{COMULISglobe}
    During the validation phase of the COMULISglobe task, a total of ten teams, including two baseline methods provided by the challenge organizers, submitted over 120 valid entries on the public leaderboard. The high number of submissions and overall strong performance demonstrated meaningful progress in aligning SHG and BF microscopy images, despite the substantial initial misalignments. In the test phase, the number of submitting teams dropped to five, with mainly the top-performing validation participants proceeding. This potentially reflects a broader issue in challenge design: teams with less competitive validation results may be discouraged from submitting to the test phase, introducing a performance bias and limiting the diversity of approaches evaluated in the final benchmark. Addressing this remains an open challenge in the design of fair and representative benchmarking competitions.
    
    One of the two baseline methods in the test phase was the affine variant of \textbf{greedy}, a fast, CPU-based deformable image registration algorithm developed by Paul Yushkevich at the Penn Image Computing and Science Lab. This variant focuses on efficient affine registration using a non-symmetric, gradient descent-based optimization, prioritizing speed over symmetry. While greedy shares conceptual roots with ANTs’ SyN \citep{avants2008symmetric}, its streamlined design and optimized computations enable rapid alignment suitable for large-scale tasks. Serving as a classical baseline, greedy provides robust reference for comparing the effectiveness of more complex registration methods in the COMULISglobe challenge task\footnote{\url{https://github.com/hansenlasse/L2R-COMULIS-baseline-greedy}}.

    The second baseline method, \textbf{globalign}, performs multimodal image alignment by globally maximizing mutual information (MI) using a novel cross-mutual information function (CMIF) computed efficiently in the frequency domain \citep{ofverstedt2022fast}. Designed primarily for low-degree-of-freedom transformations such as rigid alignment, globalign has demonstrated robust and fast performance across various multimodal biomedical image datasets, including histological images. Its algorithmic efficiency enables rapid estimation of global optima, contrasting with typical local MI maximization approaches. For this challenge, the default configuration—featuring a faster but less comprehensive search—was selected as the baseline. Adjustments to the hyperparameters may lead to improved results. The method's open-source implementation is publicly available\footnote{\url{https://github.com/MIDA-group/globalign}}.

    One of the three submissions, \textbf{VROC}, is a classical variational image registration framework that supports flexible choices of similarity measures, regularization strategies, and transformation models \citep{werner2014estimation, ehrhardt2015variational}. For this challenge task of multimodal registration of BF and SHG images, VROC combines handcrafted preprocessing with affine alignment and multi-resolution variational refinement. A custom-trained U-Net is used to synthesize SHG-like representations from BF inputs, reducing the modality gap. During inference, VROC performs extensive hyperparameter tuning, including affine step sizes, smoothness parameters, and optimization schedules, optionally preceded by a pre-registration step. The optimization uses normalized cross-correlation (NCC) as similarity metric, and incorporates a Demons-based strategy \citep{thirion1998image} for deformable refinement. This highly engineered pipeline enables robust adaptation to the modality-specific challenges of SHG/BF registration. The general VROC framework is publicly available on GitHub\footnote{\url{https://github.com/IPMI-ICNS-UKE/vroc}}.

    \textbf{next-gen-nn}, employs a hybrid pipeline that combines learning-based feature extraction with classical optimization. The feature backbone, adapted from COMIR \citep{pielawski2020comir}, is pretrained to extract modality-invariant representations, facilitating consistent feature alignment across SHG and BF inputs. Prealignment is performed using XFeat \citep{potje2024xfeat}, a sparse feature matcher optimized for multimodal pairs. During training, the mutual information between image pairs is used both as the objective function and as a criterion for selecting optimal deformation fields. Fine-tuning is performed using a similar design approach to ConvexAdam \citep{siebert2024convexadam}, a recent automatic registration framework.
    
    The \textbf{lWM} method combines classical feature matching with instance-specific deformable registration, forming a hybrid pipeline \citep{wodzinski2024automatic,wodzinski2024deeperhistreg}. The process begins with modality normalization to minimize appearance differences between BF and SHG images. Sparse keypoint correspondences are then extracted using the SuperPoint detector \citep{detone2018superpoint} and matched via the SuperGlue \citep{sarlin2020superglue} algorithm, which are both considered high-capacity learning-based matching models. To compensate for their lack of rotation and scale invariance, the method performs exhaustive matching across a range of initial rotations and scales, selecting the transformation with the highest number of valid matches. This provides a robust affine initialization. The final stage involves dense, multi-resolution deformable registration via instance optimization rather than deep learning. This choice was motivated by the limited availability of training data and the lack of pre-trained foundation models for microscopy registration, favoring a flexible, optimization-based approach with strong generalization properties.

    \begin{figure}[t!]
        \centering
        \renewcommand{\arraystretch}{0.8}
        \begin{tabular*}{\linewidth}{@{\extracolsep{\fill}} c l c c c}
        \toprule
        \# & Team/Method & TRE & TRE30 & Rank \\
        \midrule
        1 & VROC                & 2.64  & 2.86  & 0.90 \\
        2 & next\_gen\_nn       & 2.93  & 3.18  & 0.77 \\
        2 & greedy*             & 3.03  & 3.29  & 0.77 \\
        3 & lWM                 & 3.10  & 3.42  & 0.46 \\
        4 & globalign*          & 3.21  & 3.64  & 0.28 \\
        \midrule
          & Initial             & 69.6  & 102.  &      \\
        \bottomrule
        \end{tabular*}
        \includegraphics[width=1\linewidth]{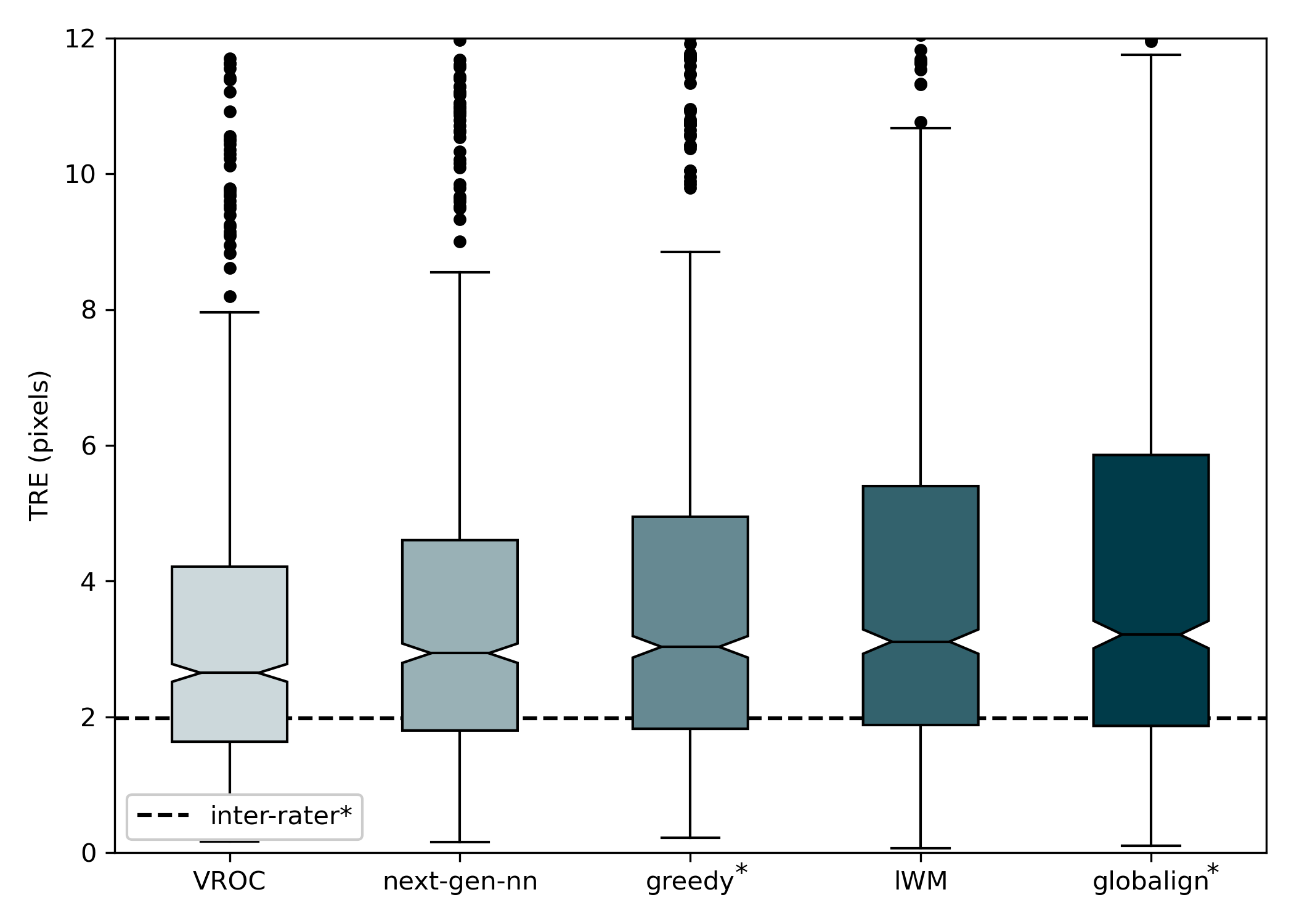}
        \caption{Results for the COMULISglobe SHG/BF task. *The mean inter-rater error was computed on a random subset of test cases. While thus not directly comparable to the evaluated methods, it provides a useful reference for the level of alignment that may be achievable. * Baseline method provided by organizers.}
        \label{results_comulis}
    \end{figure}

    All methods were evaluated on the 40 test cases using the TRE (Target Registration Error) and TRE30 (the TRE computed on the 30\% of landmarks with the largest initial errors) metrics, based on manually annotated landmarks ($\sim$25 per case). An inter-rater (three raters) error, computed on a random subset of test cases, had a mean TRE of $2.00$ pixels, serving as a rough upper bound for achievable performance. The results (cf. \ref{results_comulis}) show that all evaluated methods achieved substantial improvements over the unaligned baseline (mean TRE: 69.6). The top-ranked method, VROC, achieved the lowest mean TRE (2.64) and TRE30 (2.86), followed closely by next-gen-nn (TRE: 2.93) and greedy (TRE: 3.03), which obtained nearly identical overall ranks. lWM and globalign completed the test phase with mean TREs of 3.10 and 3.21, respectively. Despite differences in methodological paradigms, the final TRE scores among the top submissions fall within a narrow range, suggesting that both learning-based and classical pipelines can achieve similar levels of accuracy when appropriately tuned.  However, robustness remains a challenge: failure cases—defined as those with mean TRE greater than 15 pixels—occurred across all methods, with counts of 3, 1, 4, 4, and 7 for VROC, next-gen-nn, greedy, lWM, and globalign, respectively.

    The COMULISglobe SHG/BF task attracted fewer submissions compared to other Learn2Reg tasks, highlighting a potential engagement gap between the classical medical image registration and the biomedical (particularly histology) imaging communities. Notably, the majority of participants came from traditional medical image registration backgrounds, with minimal contribution from the histology and broader biomedical imaging communities, including active COMULISglobe members. Bridging these communities and fostering greater collaboration should be a priority for future editions of the challenge to better address the unique requirements of multimodal microscopy registration.

    This division was also reflected in an initially introduced second COMULISglobe task—3D-CLEM registration—based on data from \citep{krentzel2023clem} which was ultimately dropped due to its high complexity. It involved automatic registration of multimodal 3D microscopy data, specifically aligning high-resolution electron microscopy (FIB-SEM and SBF-SEM) with super-resolution fluorescence light microscopy. The task benefited from domain knowledge and presented significant challenges related to large volumes and sparse correspondences, making it less accessible to the broader registration community. We assume it would have been of great interest to microscopy specialists, a group that was underrepresented in the challenge.

    While most methods effectively solved the predominantly rigid alignment components of the task, opportunities remain for improving deformable registration performance. Robustness continues to be a key challenge, as evidenced by approximately 10\% failure cases across all methods. Addressing these limitations will be critical for the practical application of multimodal microscopy registration in biomedical research and clinical workflows.

    \subsubsection{NLST}
    The NLST task presented a highly realistic and challenging benchmark for thoracic CT registration, incorporating both lung cancer screening and radiotherapy cases. The new and extended evaluation from 2023 benefited from a substantially enhanced dataset, with over 100 landmarks per case annotated independently by two medical experts, offering a rigorous basis for assessment. The top-performing method, \textbf{LapIRNv2} \citep{mok2021conditional}, achieved the best overall rank (0.800), closely followed by the organizer-provided baseline (\textbf{corrfield} \citep{heinrich2015estimating}) (rank: 0.670) and \textbf{Birmingham} \citep{10.1007/978-3-031-21014-3_16,jia2023fourier} (rank: 0.553). These methods demonstrated low target registration errors (TRE: 1.425–1.63 mm) and excellent robustness on difficult cases (TRE30: 1.923–2.200 mm). While deep learning approaches dominated the top ranks, classical keypoint-based strategies remained competitive. Notably, the lightweight learning-based MONAI baseline\footnote{\url{https://github.com/Project-MONAI/tutorials/blob/main/3d_registration/learn2reg_nlst_paired_lung_ct.ipynb}} also performed well (rank: 0.452), highlighting the potential of efficient architectures. While teams such as UKE and UCLA reported higher TREs, all submissions significantly outperformed the unregistered baseline (initial TRE: 10.40 mm). Importantly, lung nodule TRE, a clinically critical metric, was included in the final ranking and aligned well with landmark-based scores (here the Top-4 had roughly 50\% lower errors than MONAI and UCLA with TRE$_{nodules}$: 1.135–1.434 mm), supporting its continued use in longitudinal studies. All leading methods executed within clinically acceptable runtimes of several seconds, and the task's realistic anatomical variability underlined the importance of robustness and generalizability in clinical deployment.
    
    \begin{table}[ht]
    \centering
    \renewcommand{\arraystretch}{0.8}
    \begin{tabular*}{\linewidth}{@{\extracolsep{\fill}} c l c c c}
    \toprule
    \# & Team & TRE & SDLogJacDet & Rank \\
    \midrule
    1 & LapIRNv2    & 1.441  & 0.042 & 0.800 \\
    2 & corrfield   & 1.638  & 0.038 & 0.670 \\
    3 & Birmingham  & 1.425  & 0.045 & 0.553 \\
    4 & UKE         & 2.435  & 0.062 & 0.489 \\
    5 & MONAI       & 2.042  & 0.039 & 0.452 \\
    6 & UCLA        & 2.934  & 0.071 & 0.352 \\
    \midrule
      & initial     & 10.40 &     &  \\
    \bottomrule
    \end{tabular*}
    \caption{Summary of NLST 2023 Results. TRE is the mean target registration error on anatomical landmarks in millimeters.}
    \label{tab:nlst_summary}
    \end{table}

\subsection{Learn2Reg 2024}
    The 2024 edition of the Learn2Reg challenge demonstrated active engagement from a wide range of research teams across all three new tasks. Notably, two teams—VROC and next-gen-nn—submitted algorithms consistently across all three challenge tracks, demonstrating the versatility of their methods and enabling valuable cross-task insights. Both teams shared second place on the ReMIND2Reg task. In the LUMIR challenge, next-gen-nn also secured second place, while VROC did not place among the top ten. In the COMULISglobe SHG/BF challenge, VROC achieved first place, followed by next-gen-nn in second place. These placements are based solely on participant submissions and do not include organizer-provided baselines. While no official overall ranking was announced, the consistent high performance of next-gen-nn across all tasks suggests strong generalizability of their approach. The two teams differed notably in their methodological strategies. VROC employed a consistent, predominantly optimization-based pipeline across all tasks. Their approach combined task-specific preprocessing (e.g., Gaussian smoothing, modality mapping) with affine or rigid registration using modality-adapted similarity metrics such as NCC, NGF, or MSE. For deformable alignment, they incorporated Demons-like forces and relied on extensive case-wise hyperparameter tuning, guided by a surrogate loss based on NCC. In contrast, next-gen-nn adopted a more adaptive strategy depending on the task: for ReMIND2Reg, they chose a fully optimization-based approach, while for COMULISglobe SHG/BF, they applied optimization-based fine-tuning on top of deep learning results.

    Several general findings emerged from the 2024 edition of the Learn2Reg challenge. While participation was generally strong and engagement high, most teams focused their final test submissions on the LUMIR task, a brain MRI registration challenge. This mirrors trends from previous editions, where tasks using the OASIS data also attracted the majority of participants. The relatively lower participation in more complex tasks like ReMIND2Reg, characterized by multi-modality and varying fields of view, suggests these remain challenging and potentially less accessible. In contrast, the LUMIR task yielded consistently good results across nearly all submissions, indicating it may be more motivating or tractable for participants. The COMULISglobe task—focusing on registration of histology images—attracted fewer teams. This may be due to Learn2Reg’s traditional emphasis on classical medical image registration, making biomedical image registration a more niche and specialized subfield within the challenge.

    The 2024 challenge also highlighted the absence of a standardized, robust registration framework analogous to nnUNet in segmentation. ConvexAdam \citep{siebert2024convexadam} represents a step toward this goal and was adopted by next-gen-nn in several tasks, highlighting promising research avenues focused on adaptable and reusable registration frameworks.
    
    Baseline methods, particularly classical optimization-based tools, continue to perform strongly in some tasks. For instance, NiftyReg \citep{modat2014global} achieved the best results on ReMIND2Reg, while Greedy \citep{greedy} shared second place with next-gen-nn on the COMULISglobe SHG/BF task. The LUMIR challenge was an exception, where classical baselines such as ANTs SyN were outperformed by deep learning-based methods. Notably, LUMIR had the largest dataset of all three tasks, with over 4,000 scans, which likely highlights the connection between dataset size and the success of deep learning approaches. Overall, while deep learning approaches demonstrate strong performance and continue to advance the field, the results suggest a degree of stagnation in optimization-based image registration and deep learning methods suited for smaller datasets, emphasizing the need for novel approaches that can generalize robustly across diverse clinical scenarios.

\section{Conclusion and Outlook}
    The Learn2Reg 2024 challenge provided valuable insights into the current state of medical image registration through its three distinct tasks: ReMIND2Reg, LUMIR, and COMULISglobe SHG/BF. These tasks addressed a range of clinically relevant scenarios, including intraoperative ultrasound-to-MRI alignment in brain tumor surgery, large-scale unsupervised brain MRI registration, and multimodal histopathology image analysis. The outcomes highlighted significant progress in various methodologies, from advanced deep learning approaches to refined classical optimization techniques.

    A key observation from Learn2Reg 2024 was the influence of dataset characteristics on method performance. For instance, the LUMIR task, benefiting from an extensive dataset of over 4,000 brain MRI scans, generally favored deep learning methods. Conversely, in tasks with more constrained data or unique challenges, such as the multi-modal and varying field-of-view aspects of ReMIND2Reg, or the specialized nature of COMULISglobe in histology, classical optimization-based methods like NiftyReg and Greedy demonstrated strong performance and even achieved leading results in some instances. This variability underscores the ongoing need for flexible registration techniques that can perform effectively across different data scales and modality combinations.
    
    Despite the notable advancements, several limitations and areas for further development were identified. The field still lacks a widely applicable and robust registration framework that can adapt across diverse clinical scenarios and imaging modalities without extensive task-specific adjustments. While certain frameworks show promise in adaptability, consistency in performance remains a challenge. Additionally, the differing levels of participation across tasks, particularly the lower engagement in more specialized challenges like COMULISglobe, suggests a need for better integration and outreach to broader biomedical imaging communities. Building stronger connections between traditional medical image registration and other fields, such as histology, is crucial for addressing a wider spectrum of clinical needs. The occurrence of failure cases, even among top-performing methods, also points to the persistent need for enhanced robustness in real-world applications, aiming for algorithms that are more resilient to anomalies and noise.
    
    Moving forward, Learn2Reg plans to build on these findings. Future editions will aim to broaden the scope of biomedical imaging tasks, actively seeking to involve new communities and incorporate novel modalities to address a wider range of registration problems. A strong focus will be placed on developing more generalizable and robust registration frameworks that require less manual tuning and can adapt more seamlessly to new data distributions and imaging conditions. We will also continue to refine our challenge design to encourage broader participation and foster interdisciplinary collaboration, ensuring Learn2Reg remains a valuable platform for advancing medical image registration research and its practical applications.


\acks{
The organization of the COMULISglobe challenge was made possible through the support and funding from the Chan Zuckerberg Initiative (\url{https://chanzuckerberg.com/}) under the project \textit{Advancing Imaging Through Collaborative Projects}. Their support has been instrumental in advancing the field of multimodal imaging and in facilitating the dissemination of this work (\url{https://www.comulis.eu}).

We thank Kevin Eliceiri, Bin Li, and Adib Keikhosravi for providing the publicly available multimodal biomedical dataset~\citep{eliceiri2021dataset}, which was used for the COMULISglobe SHG-BF registration task. Their contribution was instrumental in enabling the design and implementation of this multimodal image registration challenge.

We also thank Marie-Charlotte Domart, Lucy Collinson, and Martin Jones from the Electron Microscopy Science Technology Platform at The Francis Crick Institute (London, UK) for providing the dataset used in the 3D-CLEM registration task. The data were made available via the EMPIAR and BioImage Archive platforms, and their support is gratefully acknowledged.

We acknowledge the use of large language models (LLMs) to support the writing and editing of this manuscript. All content was reviewed and finalised by the authors.
}

%
\ethics{The work follows appropriate ethical standards in conducting research and writing the manuscript, following all applicable laws and regulations regarding treatment of animals or human subjects.}

\coi{We declare we do not have any conflicts of interest.}

\data{All data of this challenge paper, except for the hidden test data, are publicly available through the Learn2Reg 2024 challenge website at \url{https://learn2reg.grand-challenge.org/learn2reg-2024/}.}

\bibliography{sample}

\end{document}